\documentclass[apjl]{emulateapj}

\newcommand {\bc}{\begin {center}}
\newcommand {\ec}{\end {center}}
\newcommand {\be}{\begin {equation}}
\newcommand {\ee}{\end {equation}}
\newcommand {\eqref}[1]{equation (\ref{#1})}
\newcommand {\exref}[1]{(\ref{#1})}
\newcommand {\disp}{\displaystyle}

\shorttitle{Relation between gas density and velocity power spectra}
\shortauthors{Zhuravleva et al.}

\begin{document}
\title{The relation between gas density and velocity power spectra in galaxy clusters: qualitative
treatment and cosmological simulations}
\author{I.~Zhuravleva\altaffilmark{1,2}, E.~M.~Churazov\altaffilmark{3,4}, A.~A.~Schekochihin\altaffilmark{5,6}, E.~T.~Lau\altaffilmark{7,8}, D.~Nagai\altaffilmark{7,8,9},  M.~Gaspari\altaffilmark{3}, S.~W.~Allen\altaffilmark{1,2,10}, K. Nelson\altaffilmark{9}, I.~J.~Parrish\altaffilmark{11}}
\affil{$^1$Kavli Institute for Particle Astrophysics and Cosmology, Stanford University, 452 Lomita Mall, Stanford, CA 94305, USA
}
\affil{$^2$Department of Physics, Stanford University, 382 Via Pueblo Mall, Stanford, CA  94305-4060, USA}
\affil{$^3$Max
Planck Institute for Astrophysics, Karl-Schwarzschild-Strasse 1, 85741 Garching, Germany
}
\affil{$^4$Space Research Institute (IKI), Profsouznaya 84/32, Moscow 117997, Russia
}
\affil{$^5$The Rudolf Peierls Centre for Theoretical Physics, University of Oxford, Oxford OX1 3NP, UK
}
\affil{$^6$Merton College, Oxford OX1 4JD, UK
}
\affil{$^7$ Department of Physics, Yale University, New Haven, CT 06520, USA
}
\affil{$^8$ Yale Center for Astronomy \& Astrophysics, Yale University, New Haven, CT 06520, USA
}
\affil{$^9$ Department of Astronomy, Yale University, New Haven, CT 06520, USA;
}
\affil{$^{10}$ SLAC National Accelerator Laboratory, 2575 Sand Hill Road, Menlo Park, CA 94025, USA
}
\affil{$^{11}$ Canadian Institute for Theoretical Astrophysics, 60 St. George Street, University of Toronto, Toronto, ON M5S 3H8, Canada
}

\begin{abstract}
We address the problem of evaluating the power spectrum of the velocity
field of the ICM using only information on the plasma density fluctuations,
which can be measured today by {\it Chandra} and {\it XMM-Newton}
observatories. We argue that for relaxed 
clusters there is a linear relation between the rms density
and velocity fluctuations across a range of scales, from the largest ones, where motions are dominated by buoyancy, down to small, turbulent scales: $(\delta\rho_k/\rho)^2 = \eta_1^2 (V_{1,k}/c_s)^2$, 
where $\delta\rho_k/\rho$ is the spectral amplitude of the density perturbations
at wave number $k$, $V_{1,k}^2=V_k^2/3$ is the mean square component of the velocity field, 
$c_s$ is the sound speed, and $\eta_1$ is a dimensionless
constant of order unity. Using cosmological simulations of  relaxed galaxy
clusters, we calibrate this relation and find $\eta_1\approx 1 \pm 0.3$. 
We argue that this value is set at large scales 
by buoyancy physics, while at small scales the density 
and velocity power spectra are proportional because the 
former are a passive scalar advected by the latter. This opens an interesting
possibility to use gas density power spectra as a proxy for the
velocity power spectra in relaxed clusters, across a wide range of scales.
\end{abstract}
\keywords{galaxies: clusters: intracluster medium---hydrodynamics---methods: analytical---methods: numerical---plasmas---turbulence}

\section{Introduction}
Spectacular data accumulated by X-ray observatories on
the nearest X-ray brightest clusters of galaxies allow us to probe inhomogeneities in the
intracluster medium (ICM) over a broad range of spatial scales. These clusters typically show $\sim 5-10$\%
 density fluctuations on scales from a few tens to a few
hundred kpc \citep[][see also Schuecker et al. 2004 for earlier work]{Chu12,San12}. At the same time, the dynamics of the ICM remain largely unknown. For relaxed clusters, numerical simulations predict predominantly subsonic motions of the ICM 
on scales from $\sim$Mpc down
to a few tens of kpc with approximately Kolmogorov power spectra
(PS) of the velocity field \citep[see, e.g.,][]{Nor99,Dol05,Nag07b,Iap11,Vaz11}. 

The relatively low energy resolution ($\sim 130-150$ eV) of current X-ray CCD-type
detectors precludes accurate measurements of gas velocities 
\citep[see, e.g.,][]{Zhu13b,Tam14}. The gain in resolution can be achieved in cool cores of clusters by using grating spectrometers. Such observations provide mostly upper limits on the gas velocities $\sim$ a few hundred km/s \citep[e.g.,][]{Wer09,San13,Chu04}. One can also use Faraday
Rotation measurements to probe the ICM turbulence indirectly \citep[e.g.,][]{Vog03}.

The future Japanese-US X-ray observatory {\it Astro-H} \citep[see][launch in 2015]{2010SPIE.7732E..27T}
 should provide high-resolution ($\sim 4-7$ eV) X-ray spectra,
 allowing one for the first time to measure gas velocities directly. However,
 it will not be trivial to extract the PS of the velocity field
 \citep[see methods developed in][]{Zhu12}. The full power of the methods can only be used once the next
 generation of X-ray observatories, such as {\it
   SMART-X}\footnote{http://smart-x.cfa.harvard.edu/index.html} and
 {\it Athena+}\footnote{http://athena2.irap.omp.eu/}, are operating.

In the meantime, are there ways to probe the velocity PS with existing and near-term data? 
In this Letter, we argue that, for
subsonic motions in relaxed clusters, there is a linear relation
between the PS of density fluctuations derived from X-ray
images, and the velocity PS. Using analytical description of a passive scalar advected by fluid motions in stratified medium, we show that the linearity holds
from large scales, where motions are dominated by buoyancy,
down to small, turbulent scales, with the same 
coefficient of proportionality. In turbulent regime linear dependence was found in simulations of massive cluster with solenoidal forcing in \citet{Gas13}. It is interesting that similar situations arise in
the context of solar wind, Earth atmosphere, and the ISM
\citep[see, e.g.][]{1995ApJ...443..209A}. 

\citet{Chu12} list the following contributions to
measured density variations in clusters: (1) perturbations of the
gravitational potential; (2) deviations from the oversimplified model
profiles; (3) entropy fluctuations caused by infalling 
low-entropy gas or by gas advection; (4) pressure variations
associated with gas motions and sound waves; (5) metallicity
variations; (6) the presence of non-thermal and spatially variable
components. Cosmological simulations of relaxed clusters (Section 3), which include effects
(1)--(4), illustrate that predicted linearity holds approximately in the
case of  “natural” cosmological driving. In the companion paper
(Gaspari et al., 2014, hereafter G14), high-resolution simulations 
in a static gravitational potential with solenoidal forcing of turbulence 
are used to investigate items (3) and (4) and 
the role of isotropic thermal conduction.

\section{Velocity field and density fluctuations}
\label{sec:theory}
Let us consider slow, subsonic gas motions in a cluster
 potential. Two different regimes can be distinguished: (i) a
large-scale limit, where the dynamics are governed by
buoyancy and (ii) a turbulent regime at small scales,
where the eddy turnover time is shorter than the characteristic
buoyancy time scale. 
Below, we argue that there is a linear relation with the same constant 
of proportionality between the amplitudes of 
the gas density and velocity fluctuations in both regimes.

\subsection{Buoyancy-dominated regime (large scales)}

Assuming that the ICM can be described by standard hydrodynamics (or magnetohydrodynamics), 
the entropy $s=P/\rho^\gamma$ (where $P$ is pressure, $\rho$ density and $\gamma=5/3$ the 
adiabatic index) satisfies 
\be
\frac{\partial s}{\partial t} + {\bf V}\cdot\nabla s = 0, 
\ee
where ${\bf V}$ is the flow velocity and we have neglected any heat fluxes, 
heating or cooling of the ICM. In a static equilibrium, the entropy 
has a radial profile $s_0(r)$ (a stratified atmosphere in a gravitational well). 
As the ICM is turbulent, this 
profile will be perturbed on scales that are smaller than 
the scale height $H_s=(d\ln s_0/dr)^{-1}$ --- and if we assume that the ICM 
motions are subsonic, $V\ll c_s=\sqrt{\gamma P_0/\rho_0}$, these perturbations 
will be small: $s=s_0+\delta s$, $\delta s/s_0\ll1$. They satisfy
\be
\left(\frac{\partial}{\partial t} + {\bf V}\cdot\nabla\right)\frac{\delta s}{s_0} 
= - \frac{V_r}{H_s},
\label{ds_eq}
\ee
where $V_r$ is the radial velocity perturbation. 

If all perturbations, including $V/c_s$, 
were infinitesimal, the restoring buoyancy force on a gas 
element displaced in the radial direction would result in oscillatory 
motions --- gravity waves, or g-modes. Their frequency is
\be
\omega = \frac{k_\perp}{k}\,N,\quad
N = \sqrt{\frac{g}{\gamma H_s}} = \frac{c_s}{\gamma\sqrt{H_sH_p}}, 
\ee
where $k$ is the wave number of the oscillations, 
$k_\perp$ its projection perpendicular to the radial direction,  
$N$ is the Brunt-V$\ddot{\rm a}$is$\ddot{\rm a}$l$\ddot{\rm a}$ frequency, 
$g$ is the acceleration of gravity, $H_p=(d\ln P_0/dr)^{-1}$ is the pressure 
scale height,\footnote{In an isothermal cluster, $H_p/H_s = \gamma-1 = 2/3$. In simulated clusters (Section 3) typical values of $H_p$ and $H_s$ are $\sim 200-300$ kpc at distance 100 kpc from the center.}   
and the last equality follows from the hydrostatic force balance, $dP_0/dr = \rho_0 g$. 
The density perturbations associated with these motions are
\be
\left(\frac{\delta\rho}{\rho}\right)^2 = 
\left(\frac{1}{\gamma}\frac{\delta s}{s}\right)^2 = 
\left(\frac{V_r}{\gamma\omega H_s}\right)^2
= \frac{H_p}{H_s}\frac{k^2}{k_\perp^2}\left(\frac{V_r}{c_s}\right)^2,
\label{rhoV_lin}
\ee
which follows from \eqref{ds_eq} if the advection term is neglected;  
we have suppressed $0$'s in the subscripts of equilibrium quantities. 
The relationship between $\delta\rho/\rho$ and $\delta s/s$ is a consequence of local pressure balance 
($\delta P/P\ll\delta\rho/\rho$), which holds for subsonic motions. 

In reality, perturbations are not infinitesimal and the question is 
to what extent the linear relationship between density and velocity 
survives in the strongly nonlinear regime, 
when the advection term in \eqref{ds_eq} is not negligible. 
The argument that this relationship does survive 
depends somewhat nontrivially on the strength of the ICM turbulence 
at the outer (energy-injection) scale. The key parameter is the Froude 
number, 
\be
{\rm Fr} = \frac{V_{\rm rms}}{L_\perp N} =  {\rm Ma}\,\frac{\gamma\sqrt{H_s H_p}}{L_\perp},
\ee
the ratio of the nonlinear 
decorrelation and linear Brunt-V$\ddot{\rm a}$is$\ddot{\rm a}$l$\ddot{\rm a}$ 
frequencies at the outer scale ($V_{\rm rms}$ is the rms velocity of the turbulent motions, 
${\rm Ma}=V_{\rm rms}/c_s$ is the Mach number and $L_\perp$ is the outer scale
perpendicular to the radial direction).

If ${\rm Fr}\ll 1$, the turbulence will tend to a stratified, anisotropic 
regime, in which $k_\perp\ll k\approx k_r$ and the gravity-wave frequency 
$\omega=Nk_\perp/k\ll N$ stays comparable to the nonlinear 
decorrelation rate $k_\perp V_\perp$ \citep[a type of ``critically balanced'' state; see][]{Lin06,Naz11}. 
In this regime, buoyancy remains important scale by scale ($k$ by $k$) 
as the energy cascades to smaller scales (larger $k$), 
the relationship \exref{rhoV_lin} is approximately satisfied,  
and the typical velocity of the motions is dominated by the 
perpendicular velocity: $V\sim V_\perp\sim (k_r/k_\perp)V_r\gg V_r$ 
(by incompressibility). Therefore, we deduce, at each $k$, 
\be
\left(\frac{\delta\rho_k}{\rho}\right)^2
= \eta^2\left(\frac{V_k}{c_s}\right)^2,
\label{rhoV_turb}
\ee
where $\eta$ is a scale-independent dimensional constant of order unity. 
Here $\delta\rho_k$ and $V_k$ are some suitably defined fluctuation amplitudes 
at scale $k^{-1}$ (see Section 3.2), rather than Fourier components, 
they are related to the 3D spectrum $E_k$ by $|V_k|^2\sim kE_k$. 

For this kind of turbulence, it is possible to show \citep[see][and references therein]{Lin06,Naz11}
that the energy spectrum of the perpendicular motions (and, therefore, of the density perturbations) 
is Kolmogorov in the perpendicular direction, $E_k\sim \varepsilon^{2/3}k_\perp^{-5/3}$, 
where $\varepsilon$ is the energy flux, and much steeper in the radial direction, 
$E_k \sim N^2k_r^{-3}$ \citep{Dew97,Bil01}, 
but that as the turbulent cascade proceeds to smaller scales, turbulence 
becomes less anisotropic, eventually reaching isotropy ($k_\perp\sim k_r$, 
$V_\perp\sim V_r$) at the so-called Ozmidov scale, $k\sim k_{\rm O}=N^{3/2}\varepsilon^{-1/2}$, 
where the perpendicular and radial spectra meet and 
the ``local'' Froude number is $V_{k_{\rm O}}k_{\rm O}/N\sim 1$ 
(\citealt{Ozm92}; for the latest numerical results, see \citealt{Aug12}). 
Beyond this scale ($k>k_{\rm O}$), the stratified cascade turns into the usual 
isotropic Kolmogorov cascade, in which $k_\perp\sim k_r$, $V_r\sim V_\perp$, 
$E_k\sim \varepsilon^{2/3}k^{-5/3}$,  
the nonlinear decorrelation rate becomes dominant compared to the 
linear gravity-wave frequency $\omega\sim N$ --- and the 
relation \exref{rhoV_turb} continues to be satisfied, but for 
reasons unrelated to the buoyancy physics and considered in Section \ref{small}. 
 
If ${\rm Fr}\sim 1$ at the outer scale, the turbulence can be assumed 
isotropic, with $k_\perp\sim k_r\sim k$ and $V_r\sim V_\perp\sim V_{\rm rms}$
already at $kL_\perp\sim 1$, and so the relation \exref{rhoV_turb} is satisfied 
at the outer scale. Again, it will continue to be 
satisfied at $kL > 1$, as we are about to explain.
This is probably the more common 
situation: in relaxed clusters (see Table 1), 
${\rm Fr}\sim 0.3 \left (L_\perp/300~{\rm kpc} \right )^{-1}\sim 0.3-1$.

Thus, in the entire interval of scales where buoyancy matters 
(which may or may not extend beyond the outer scale, depending on 
the value of ${\rm Fr}$), one can expect a linear relation between the 
velocity and the density perturbation with a coefficient of order unity, 
\eqref{rhoV_turb}.

\subsection{Turbulent regime (small scales)}\label{small}
Consider now the limit of small scales, $k > k_{\rm O}$. 
At these scales (the ``inertial range''), the nonlinear decorrelation (eddy turnover) rate 
becomes increasingly (with $k$) greater than the Brunt-V$\ddot{\rm a}$is$\ddot{\rm a}$l$\ddot{\rm a}$
frequency. This means that the right-hand side of \eqref{ds_eq} can 
be neglected and so the entropy fluctuations are a passive 
scalar field. Since $\delta\rho/\rho = -(1/\gamma)\delta s/s$, so are 
the density fluctuations. 

A scaling theory for such passive fluctuations in the inertial range 
can be constructed along the lines of the Obukhov--Corrsin theory \citep{Obu49,Cor51}.  
The total variance of $\delta\rho/\rho$ is conserved, and so there 
should be a constant flux $\varepsilon_\rho$ of it towards smaller scales: 
\be
\left(\frac{\delta\rho_k}{\rho}\right)^2 \sim \frac{\varepsilon_\rho\tau_k}{c_s^2},
\label{const_flux}
\ee
where $\varepsilon_\rho$ is the scalar dissipation rate, $\tau_k$ is the typical ``cascade time'' and $c_s^2$ appears in the right-hand 
side simply to ensure convenient normalization of the flux $\varepsilon_\rho$ to energy units. 
Similarly, for the turbulent motions themselves, 
\be
V_k^2 \sim \varepsilon_V\tau_k,
\label{K41}
\ee
where $\varepsilon_V$ is the flux of the kinetic energy and $\tau_k$ is the same 
cascade time as in \eqref{const_flux} because both $\delta\rho/\rho$ and ${\bf V}$ 
are mixed by the same velocity field. From \eqref{K41}, $\tau_k\sim V_k^2/\varepsilon$
and, therefore, from \eqref{const_flux}, 
\be
\left(\frac{\delta\rho_k}{\rho}\right)^2 \sim \frac{\varepsilon_\rho}{\varepsilon_V}
\left(\frac{V_k}{c_s}\right)^2.
\label{rhoV_ir}
\ee
Thus, the spectrum of the passive scalar follows the spectrum of the velocity 
field in the inertial range. The constant of proportionality in \eqref{rhoV_ir}, 
$\varepsilon_\rho/\varepsilon_V$, depends on how much scalar variance, compared to 
kinetic energy, arrives into the inertial range from the larger, buoyancy-dominated 
scales. Therefore, the relationship \exref{rhoV_ir} should be matched with 
\exref{rhoV_turb}: $\varepsilon_\rho/\varepsilon_V\sim\eta^2$. 

Note that these conclusions are independent of what precisely the spectrum of the 
turbulence is because the argument above was based only on the assumption that the cascade times are 
the same for the passive scalar and for the turbulence that advects it. 
This is reassuring because ICM turbulence is certainly not a simple hydrodynamic
Kolmogorov turbulence of an inertial fluid. At the very least, it is 
magnetohydrodynamic, as clusters are known to host dynamically significant 
magnetic fields \citep[see e.g.,][and references therein]{Sch06,Ens06}. 
Since $\beta=8\pi P/B^2\sim10^2\gg1$, these fields do not significantly 
modify buoyancy physics at large scales.\footnote{At least not on the qualitative level. 
Strictly speaking, one ought to worry about the instabilities caused 
by anistropic heat fluxes \citep[see, e.g.,][and references therein]{Bal00,Kun11,Qua08}.}
In the inertial range, the turbulence may be MHD rather than hydrodynamic, 
but density fluctuations continue to behave as a passive scalar in 
Alfv\'enic MHD and even kinetic turbulence \citep{Sch07,Sch09} and so the above argument 
continues to hold.\footnote{Matters may become more complicated at scales 
below the collisional mean free path, where density fluctuations 
are subject to collisionless damping \citep{Sch09}, but such small scales are unlikely 
to be observable in the near future.}
 
Thus, the conclusion from this section 
is that the relation \exref{rhoV_turb} holds with the same proportionality constant both 
at scales where buoyancy physics matters and at those where it does not 
and independently of whether ${\rm Fr}$ is small or order unity 
(i.e., independently of how vigorous the turbulence is and of whether it 
is isotropic Kolmogorov turbulence or anisotropic stratified one).   
Below, we verify these arguments using a sample of cosmological hydrodynamic 
simulations of galaxy clusters.

\begin{deluxetable}{cccc}
\tabletypesize{\footnotesize}
\tablecaption{Sample of simulated galaxy clusters}
\tablewidth{0pt}
\tablehead{
\colhead{ClusterID} &
\colhead{$\disp$ r$_{500c}$, kpc} &
\colhead{M$_{500c}$, 10$^{14}$$\cdot$M$_{\odot}$} &
\colhead{{\rm Ma}}
}
\startdata
CL149 & 814.4& 1.83&0.48\\
CL21 & 1215.2& 6.08& 0.21\\
CL223 & 824& 1.9&0.29\\
CL25 & 1095.8& 4.46&0.32\\
CL27 & 1154.5& 5.21&0.23\\
CL42 & 1297.7& 7.4&0.27
\enddata
\tablecomments{Mach number ${\rm Ma} = V_{\rm rms}/c_s$ is calculated as the RMS of the velocity 
in the central 500 kpc after subtracting the mean velocity in this region.} \\
\label{tab:sample}
\end{deluxetable}

\section{Calibration with cosmological simulations}
\label{sec:sim}
\subsection{Simulations and sample of galaxy clusters}
We use high-resolution cosmological simulations of galaxy clusters in a flat $\Lambda$CDM model with WMAP
five-year cosmological parameters: $\Omega_m = 1 -
\Omega_{\Lambda} = 0.27$, $\Omega_b = 0.0469$, $h = 0.7$ and $\sigma_8
= 0.82$. The simulations are performed
using the Adaptive Refinement Tree (ART) $N$-body+gas-dynamics code
\citep{Kra99, Kra02,Rud08}, which is an Eulerian code that uses
adaptive refinement in space and time and non-adaptive refinement in
mass \citep{Kly01} to achieve the dynamic ranges necessary to resolve the cores
of halos formed in self-consistent cosmological simulations. Details of the 
zoom-in simulations used here can be found in \citet[][Section 3.1]{Nel14}.

Our sample includes 6 relaxed clusters at $z=0$. Their X-ray
morphology exhibits spherical or elliptical symmetry with no
filamentary or clumpy substructures within $r_{500c}$
\citep[see][]{Nag07}. We analyze non-radiative (NR) runs only, because
these involve physical processes directly relevant to
those discussed in Section 2.

\begin{figure}
\centerline{\includegraphics[trim=30 0 20 0,width=0.95\linewidth]{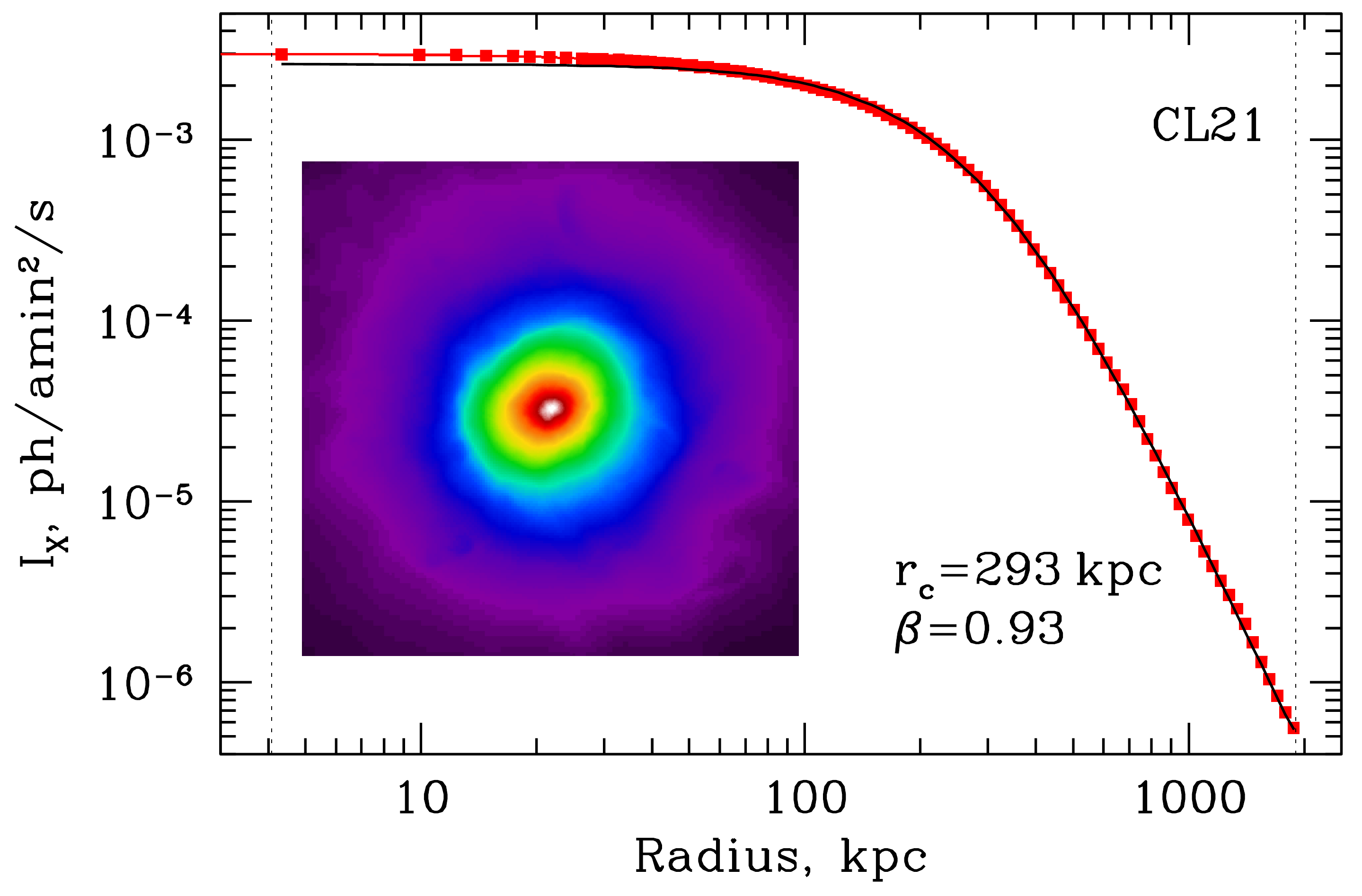}}
\caption{X-ray surface brightness (2$\times$2 Mpc) of simulated cluster CL21 (inset), 
and its radial profile (main plot, red points).
Black curve: best-fitting $\beta$ model. Vertical dash lines: the fitting interval.
\label{fig:sb}}
\end{figure}

We analyze fluctuations in the central $\sim 500$ kpc (radius)
region. The gas motions are subsonic with ${\rm Ma}\sim 0.2-0.5$ (Table \ref{tab:sample}). X-ray images
of the NR relaxed clusters do not show any prominent subhaloes or
clumpy structures within $\sim 500$ kpc (Fig.~\ref{fig:sb}, inset). Only
cluster CL25 has a small subhalo, which we remove from the
analysis. Subhaloes/clumps, which are present in 3D data and are not
visible in projection, only slightly affect our results.  E.g., for
cluster CL149 the exclusion of clumps with $\delta\rho/\rho > 1$ 
removes $\sim$1.8\% of the volume and changes the
total rms of $\delta \rho/\rho$ by $\sim$10\%. The changes in
${\rm Ma}$ are less than 1\%.

\subsection{Power spectra of density  and  velocity fluctuations}
For each cluster in our sample, we calculated the 3D emissivity-weighted
electron density as $\displaystyle n_{e,X}=n_e\sqrt{\Lambda(T)}$, where $\Lambda(T)$ is the gas emissivity \citep{Sut93}, and three components of the normalized velocity field, 
$V_{x,y,z}/c_s$. Projecting the squared density along one of the directions, 
the X-ray surface brightness (SB), $I_X(x,y)\propto\disp\int
n_{e,X}^2(x,y,z)dz$, is obtained. We calculate
the spherically-symmetric radial profile of the SB, and
approximate it with a $\beta$ model (Fig. \ref{fig:sb}). Dividing the density by the corresponding 3D
$\beta$ model and subtracting unity, the 3D density fluctuations
${\delta \rho}/{\rho}$ are obtained. The center of
each cluster is chosen as the peak of the gas density within the central $\sim 50$
kpc region and confirmed by the visual inspection.

\begin{figure}
\centerline{\includegraphics[trim=20 160 40 90,width=0.95\linewidth]{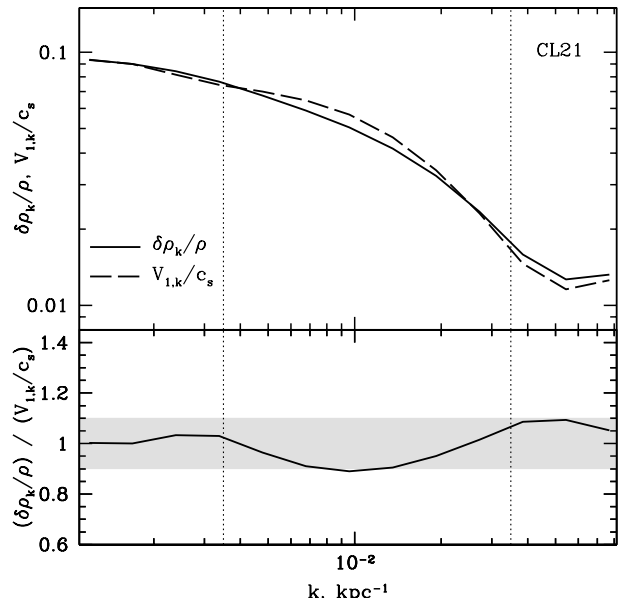}}
\caption{Amplitudes of density and velocity fluctuations (top panel)
  and their ratio (bottom panel) for CL21 cluster. A convention $k=1/\lambda$
  without a factor $2\pi$ is used throughout. The ratio is consistent with unity with a scatter $<$ 10\% (gray shaded region) at scales $\sim 30-300$ kpc (vertical dotted lines), which are least affected by numerical artifacts (see Section 3.2).} 
\label{fig:rhovel}
\end{figure}

We then use the modified $\Delta-$variance method \citep{Are12} to
calculate $P_{3D}(k)$, the PS of the 3D density fluctuations and
velocity.  These spectra are converted to fluctuations amplitudes 
$A_k=\sqrt{4\pi P_{3D}(k)k^3}$. Fig.~\ref{fig:rhovel}
shows an example of such amplitudes of the density fluctuations
$\delta\rho_k/\rho$ and of rms velocity component 
$V_{1,k}=\sqrt{(V_{x,k}^2+V_{y,k}^2+V_{z,k}^2)/3}$ for cluster CL21.
These amplitudes follow
each other over a broad range of scales. Their ratio 
$\eta_1= ({\delta \rho_k}/{\rho})/({V_{1,k}}/c_s)$ is close to unity
with $\sim 10$\% deviations.

\begin{figure}
\centerline{\includegraphics[trim=20 390 40 90,width=0.95\linewidth]{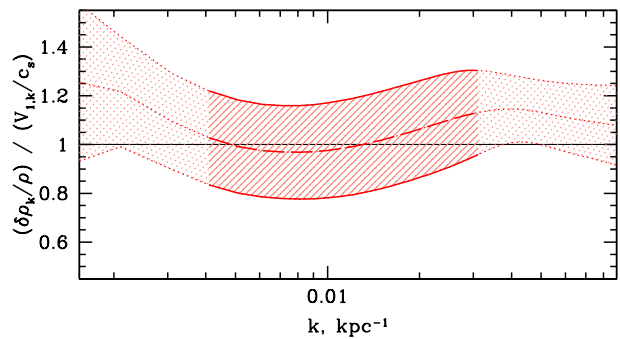}}
\caption{Sample-averaged ratio of the amplitudes of density and velocity fluctuations.
Shaded region: scatter over the sample.
Solid curves and shadows: the range of scales least affected by numerical
  artifacts (see Section 3.2). The ratio is 
$\eta_1 = 1 \pm 0.3$ at scales $\sim 30-300$ kpc$^{-1}$.}
\label{fig:rat}
\end{figure}

Even though the relationship \exref{rhoV_turb} is found to hold over a
broad range of scales, the amplitudes at the smallest and the largest
scales are affected by several artifacts. At
$k\gtrsim 4\cdot 10^{-2}$ kpc$^{-1}$, the limit of numerical resolution is
reached.  At the largest scales, the amplitude is
  (a) sensitive to the underlying model used to correct for the global
  structure of the cluster; (b) affected by uncertainties due to
  stochastic nature of perturbations. The sensitivity to (a) is
  estimated by experimenting with different underlying models
  (e.g., non-spherical $\beta$ models, averaged profiles). In order to
  evaluate the uncertainties (b), we experimented with multiple realizations of
  a Gaussian field that had a PS similar to that of
  density/velocity fluctuations in simulated clusters.  As expected
  there are large variations at $k\sim 1/L$, where $L=1000$ kpc is the
  size of the box. At $k\ge 3/L\sim 3\cdot 10^{-3}~{\rm kpc}^{-1}$ the
  variations drop down to $\sim 5-10$\%.

The ratio $\eta_1$ of the amplitudes of density and velocity fluctuations averaged
over a sample of relaxed clusters is shown in
Fig.~\ref{fig:rat}. It is consistent with $\eta_1=1$ at scales
$\sim 30-300$ kpc, with a relatively modest scatter $\lesssim$30\%. 
There are many possible reasons for this scatter. In particular, the
  presence of individual subhaloes, the choice of the cluster center and the
  underlying model, effects of AMR resolution and of finite ${\rm Ma}$
  (see G14) can all contribute to variations at this level.

In order to assess the effects of the AMR resolution on our
results, we resimulated one cluster, varying the
maximum refinement levels from 6 to 9 (the default one is 8). In the lowest-resolution runs, both density and velocity
fluctuations are suppressed compared to the high-resolution runs at
all scales, except for the largest $\sim300$ kpc.  The
amplitudes of density fluctuations in simulations with the refinement
levels 8 and 9 globally follow each other at scales $\sim$ 50-300 kpc. However, at some scales, deviations are up to a factor 1.3. The velocity amplitudes are the same at scales  $\sim 300-100$ kpc and differ by a factor 1.5 at $\sim 50$ kpc. Despite individual PS varying with the AMR resolution, the ratio of density and velocity fluctuations in all runs is still close to unity, with the scatter up to 25\% at scales $\sim 50-300$ kpc.

\section{Conclusions}
In this Letter, we have addressed the problem of constraining the gas velocity PS in relaxed galaxy clusters using the observed density fluctuations. We argue that

\begin{itemize}
\item the rms of density and velocity
  fluctuations are linearly related across a broad range of scales in
  both buoyancy-dominated and turbulent regimes; 
\item the constant of proportionality between them is set at large scales by gravity-wave physics and remains approximately the same in the non-linear turbulent regime;
\item cosmological simulations of
  relaxed clusters give a proportionality coefficient $\eta_1 \sim 1\pm 0.3$
   between the amplitude of the density fluctuations and the rms component of the flow velocity;
\end{itemize}

It is an interesting conclusion that, if the energy-injection scales are large enough
(e.g., $\sim10^2$~kpc for merger-driven turbulence), stratification leads to anisotropy 
($V_\perp\gg V_r$, $k_\perp\ll k_r$), whereas turbulence driven at small scales 
(e.g., $\sim 10$~kpc, as in the AGN-driven case) will be isotropic---these are 
the ${\rm Fr}\ll1$ and ${\rm Fr}\sim1$ cases discussed in Section 2.1. 
Indeed, in cosmological simulations, where turbulence is primarily driven by mergers, 
we see perpendicular velocities slightly larger than the radial ones in the central 500 kpc.  

Admittedly, our simulations suffer from insufficient dynamic
range and do not include all relevant physical processes. For
instance, thermal conduction could erase some of the
temperature/density fluctuations and break the relation \exref{rhoV_turb}. 
Some of these effects are considered in the
companion paper (G14), where a series of high-resolution
hydrodynamic simulations is carried out, with varying ${\rm Ma}$ 
and isotropic conductivity. 

It should be possible to verify the relation \exref{rhoV_turb} 
using future direct velocity measurements with {\it Astro-H} 
(combining with current observations). Strong deviations from $\eta\sim1$ would
suggest interesting microphysics or the dominance of 
other sources of density fluctuations.

In conclusion we have shown that the analysis of SB fluctuations in X-ray images offers a novel way to 
estimate the velocity PS in relaxed galaxy clusters. 
In general, proportionality 
between the density and velocity amplitudes for subsonic motions 
is probably a generic feature of small perturbations in stratified
atmospheres. 

\acknowledgements EC acknowledges useful discussions with Henk Spruit and Ewald Mueller. DN, EL and KN acknowledge support by NSF grant AST-1009811, NASA ATP
 grant NNX11AE07G, NASA Chandra grants GO213004B and TM4-15007X, the
 Research Corporation, and by the facilities and staff of the Yale
 University Faculty of Arts and Sciences High Performance Computing
 Center.

\bibliographystyle{apj}

\end{document}